\documentclass[twocolumn,showpacs,prl]{revtex4}

\usepackage{graphicx}%
\usepackage{dcolumn}
\usepackage{amsmath}
\usepackage{latexsym}

\begin {document}

\title
{Diffusion limited friendship network: A model for six degrees of separation}
\author
{
S. S. Manna
}
\affiliation
{
Satyendra Nath Bose National Centre for Basic Sciences
Block-JD, Sector-III, Salt Lake, Kolkata-700098, India
}
\begin{abstract}
A dynamic model of a society is studied where each person is an uncorrelated and 
non-interacting random walker. A dynamical random graph represents 
the acquaintance network of the society whose nodes are the individuals 
and links are the pairs of mutual friendships. This network exhibits 
a novel percolation like phase transition in all dimensions. On 
introducing simultaneous death and birth rates in the population we 
show that the friendship network shows the six degrees of separation 
for ever after where the precise value of the network diameter depends 
on the death/birth rate. A SIS type model of disease spreading shows that this
society remains always healthy if the population density is less than certain threshold value.
\end{abstract}
\maketitle

      Though at present the human population of the world has attained a 
   very large size, more than $6 \times 10^9$  precisely, it is believed that 
   any two randomly selected persons in the world are connected by a short 
   chain of intermediate acquaintances, typically of length 6. This 
   phenomenon is referred as the ``six degrees of separation''. The idea 
   originated from the famous letter distribution experiment of
   Milgram in 1960s \cite {Milgram}. Since then any network of $N$ nodes 
   is said to display six degrees of separation if its diameter is small
   and grows at most as $\log N$ \cite {WS, Barabasibook}. 

      Most human communications, especially the information exchanges, take
   place directly between individuals when they are at close proximity
   to one another. The spread of news, rumors, jokes and fashions all 
   take place by communications among individuals. More 
   importantly the infectious diseases also spread by person-to-person contact 
   and the structure of network of such contacts has important effects on
   the nature of the epidemics. Naturally the speed of
   spreading in general is faster for a network with small diameter.

      There are important models of the social networks like Small-world 
   network (SWN) that displays the six degrees of separation \cite {WS}.
   Also the process of spreading of epidemics is modeled by a 
   susceptible-infected-susceptible (SIS) model \cite {Anderson} in which a 
   non-equilibrium phase transition takes place from a healthy society to 
   an infected society at a critical value of the infection probability
   \cite {Anderson, Marro, Bailey, Vesp}. 

      All these models of the social networks as well as for the spreading of the infectious 
   diseases consider a static picture of the society. More precisely
   static individuals are positioned at the nodes of certain graphs and
   a person interacts with only a fixed set of neighbours determined by the degree of the
   node. Where as in actual society the number of acquaintances of a person
   increases with time. Everyday a person goes to office, market, theaters, clubs etc. and therefore
   gets acquainted with other people who were unknown to him. By the same movements
   a person becomes exposed to infections by others or transfers his own infection to
   others. Again not all the friends of an infected person has the chance of
   getting infection, only those friends who come close to this person
   has the risk of infection. In this paper we study this basic property of a dynamical society where
   individuals are not static objects but move continuously and therefore comes
   in contact with other people. To make a simple model we have considered
   the diffusive motion of the individuals and modeled the society by a
   set of random walkers. Specifically
   in our model (i) unlike static models the number of acquaintances of a person evolves with
   time (ii) irrespective of how many friends an infected
   person has, he may infect only those friends who come to his close proximity, this
   is unlike to the ordinary SIS type models. What we achieve are:
   (i) with the introduction of a death/birth rate the society indeed shows
   the six degrees of separation effect (ii) there is a threshold density of
   population, below which the society is always healthy (iii) a very interesting
   theoretical observation that the associated dynamical random graph has
   a non-trivial dimension dependent critical behaviour.

      Over last few years it is becoming increasingly evident that highly
   complex structures of many social \cite {Newman}, biological \cite {Jeong,Sole}, electronic
   communication systems \cite {web,Faloutsos} etc. can be modeled by simple graphs. 
   Erd\"os and R\'enyi 
   studied the well known random graphs (RG) of $N$ nodes where each pair of
   nodes is connected with a probability $p$ and the graph shows a
   continuous phase transition at $p_c=1/N$ \cite {Erdos}.
   Scale-free networks (SFN) are characterized by the power law decay of the 
   nodal degree distribution function: $P(k) \sim k^{-\gamma}$. Two very 
   important networks in electronic communication system like World Wide Web \cite {web}
   and the Internet \cite {Faloutsos} are observed to
   possess the scale-free property. Barab\'asi and Albert (BA) proposed a
   model for a growing SFN where nodes are linked with the preferential 
   attachment probability \cite {barabasi,albert}. Other routes, e.g., static \cite {Caldarelli}
   and quasistatic \cite {Mukherjee} models to obtain SFNs are also studied. Assigning a Hamiltonian
   correlations are studied in the optimized networks keeping biological
   networks in mind \cite {Lassig}. SWNs with random walkers capable of making
   long distance jumps are studied in \cite {Manrubia}.

      In our model each member of the society executes a simple uncorrelated and non-interacting random 
   walk on a regular lattice. Initially the population of $N$ persons are released on the 
   square lattice of size $L \times L$ at randomly selected positions.
   The system then starts evolving with time. At each time step each
   person makes a jump to one of its neighbouring lattice positions with
   equal probabilities. Each person represents a node of the growing 
   acquaintance network and a link is established between two nodes the moment 
   the corresponding pair of persons come in contact to each other at the same
   position and at the same time \cite {sublattice}. Gradually the
   number of links among the individuals grow. Thus the set of $N$ nodes and the
   set of links among these nodes define our network called
   as the Diffusion limited friendship network (DLFN) where as the associated
   graph is referred as the Dynamical random graph (DRG).

%---------------------------------------------------------------------------
\begin{figure}[top]
\begin{center}
\includegraphics[width=6.5cm]{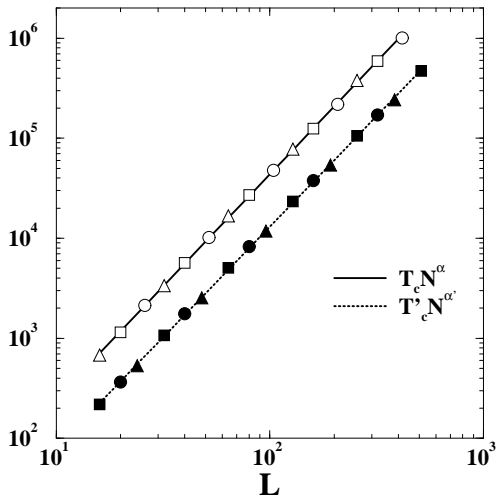}
\end{center}
\caption{
Plot of the scaled characteristic times ${\cal T}_c$ (opaque symbols) and 
${\cal T'}_c$ (filled symbols) vs. the system size $L$ for three different
populations: $N$ = 32 (circle), 64 (square) and 128 (triangle). 
}
\end{figure}
%---------------------------------------------------------------------------

   All persons which are at the same lattice site immediately become friends and 
   the associated subgraph with these people become a clique.
   At each time many such cliques are formed at different sites.
   All these cliques remain for ever, they never get destroyed, moreover they
   grow in sizes as time proceeds. At the early times, the 
   number of links is small and the DRG has many different isolated components
   of different sizes. The size of a component is determined by its number of nodes
   and the giant component has the largest size. The giant component
   not only grows by including new nodes into it but also by the process of merging equally large components.
   After some slow initial growth the giant component grows very fast and
   its size become proportional to $N$. This behaviour is just like the threshold phenomenon
   in a continuous phase transition e.g., what happens in a random graph \cite {Erdos}.
   The whole DRG ultimately reaches the 
   limiting stage of a giant $N$-clique when each node is linked to all other nodes.

%---------------------------------------------------------------------------
\begin{figure}[top]
\begin{center}
\includegraphics[width=6.5cm]{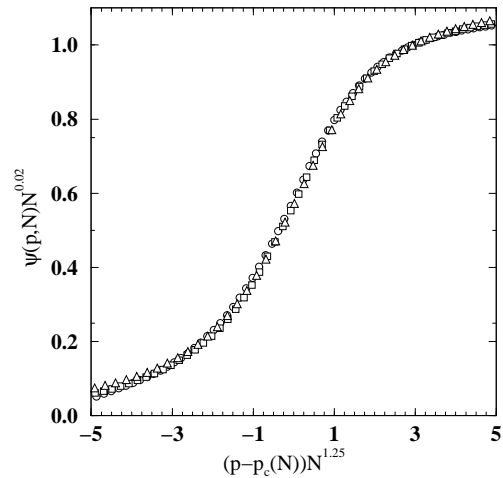}
\end{center}
\caption{
The scaling of the order parameter $\psi(p,N)$ at the critical region $p_c(N)$
in the DRG: N=64 (circle), 128 (square) and 256 (triangle).  
}
\end{figure}
%---------------------------------------------------------------------------

      We first characterize DRG to compare with RG. DRG has
   two characteristic time scales. ${\cal T}_c$ measures
   the time required for the phase transition and is observed to vary like
   $L^z/N^{\alpha}$. In a mean-field limit when the
   density $\rho=N/L^d$ is small this variation is estimated in the following way:
   If a person randomly walks a linear distance $R$ in $d$-dimension in time ${\cal T}_c$ 
   then $R \sim {\cal T}^{\mu}_c$ and therefore around $L^d/R^d$ such $d$-dimensional spheres
   are needed to cover the volume of size $L$ which is $N$ itself. This gives
   ${\cal T}_c \sim L^{1/\mu}/N^{1/\mu d}$ i.e., $z=1/\mu$ and $\alpha=1/\mu d$ in general
   and therefore $z=2$ and $\alpha=1$ for ordinary random walks ($\mu=1/2$) in two dimension.
   Fig. 1 shows the scaled plot of ${\cal T}_c$ for different $L$ and $N$ values and a good 
   collapse of the data is observed but for $\alpha \approx 0.89$ and $z \approx 2.25$.
   We believe the difference in the exponents from the mean-field values 
   are due to finite size of the system.
   At a second characteristic time ${\cal T'}_c  \sim L^{z'}/N^{\alpha'} $ the DRG becomes an $N$-clique
   where $z'$ and $\alpha'$ are estimated as 2.22 and -0.33 respectively for $\mu=1/2$ and $d$=2. The positive 
   value of $\alpha$ and the negative value of $\alpha'$ are consistent with intuition:
   for a fixed $L$ but with increasing $N$, less number of steps per person are necessary
   for the giant component to include all nodes, but a larger number of steps are
   required to form the $N$-clique. The values of $z$ and $z'$ are likely to be the same.

      The link density $p(t,N)$ at a time $t$ for an $N$ node network
   is defined as the ratio of the number of links to its maximum 
   possible number $N(N-1)/2$. Numerically we find the following scaling form:
   \begin {equation}
   p(t,L) \sim {\cal F} (t/L^z)
   \end {equation}
   where the scaling function ${\cal F}(x) \sim x^{\alpha}$ and $\alpha$ and $z$ are approximately
   found to be 0.89 and 2.25 again.

      The order parameter $\psi(p,N)$ of this transition is the average fraction 
   of nodes in the giant component for a link density $p$. The critical link 
   density at the transition point $p_c$ is defined by
   $\psi(p_c,N)=1/2$ and is observed to vary with $N$ as:
   $p_c = b/N^{\alpha}$ with $b \approx 1.28$ and $\alpha \approx 0.89$ as before.
   As the mean-field calculation gave $\alpha=1/\mu d$, we see that only
   the ordinary random walks in two dimensions with $p_c =1/N$ correspond to
   the random graphs \cite {Erdos} but for other walks with different $\mu$
   and in different dimensions $p_c(N)$ have non-trivial dimension dependence.
   A scaling plot for the order parameter is shown in Fig. 2 where we plot 
   $\psi(p,N)N^{-\beta/\nu}$ vs. $[p-p_c]N^{1/\nu}$. An excellent collapse of 
   the data shows that the order parameter has the following scaling form:
 
\begin {equation}
   \psi(p,N) \sim N^{\beta/\nu}{\cal G}[(p-p_c)N^{1/\nu}]
\end {equation}
   with $\nu \approx 0.8$ and $\beta \approx 0.02$.

%---------------------------------------------------------------------------
\begin{figure}[top]
\begin{center}
\includegraphics[width=8.0cm]{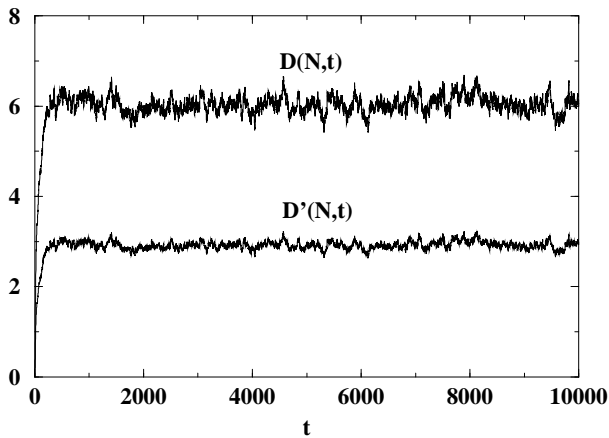}
\end{center}
\caption{
The fluctuating diameters ${\cal D}(N,t)$ and ${\cal D}'(N,t)$ of the DLFN ($q$=0.43, $N$=64, $L$=64)
plotted in the upper and the lower curves with time. The data is averaged 
over 100 configurations and have steady averages of $6.00 \pm 0.05$ and 
$2.87 \pm 0.05$. 
}
\end{figure}
%---------------------------------------------------------------------------

      The topological distance between a pair of nodes is the number 
   of links on the shortest path connecting them and the diameter is 
   the maximum of such paths. The average diameter ${\cal D}(N)$ is 
   measured over many independent configurations. The configuration 
   average of the mean distance between an arbitrary pair of nodes is
   denoted by ${\cal D}'(N)$. As the system evolves, both measures
   first increase with time, reach their maxima and then decrease very slowly,
   finally saturates to a fixed value for a long time. The maximum of the diameters
   occur at the characteristic times ${\cal T}_c$. As expected the nodal degree distribution 
   of the giant component at the transition point is a Poisson distribution similar
   to RG, since there is no preferential link attachment probability in this model as in the
   scale-free networks.

      The network described so far has a major drawback that it assumes each
   individual as immortal. As a consequence the DRG becomes an $N$-clique at time ${\cal T'}_c$.
   To make our model more realistic, we therefore introduce a probability of death and birth in the population
   but with equal rates to keep the population conserved.
   More precisely at each time step only one randomly selected individual is killed with 
   a probability $q$. As a consequence all links associated with the node
   representing this individual are immediately deleted. This may result the
   fragmentation of the particular component of the dynamical graph which belonged this node.
   A fresh determination of the different components of the DLFN
   especially the giant component is done immediately
   before the system proceeds to the next time step. At the same time we assume 
   that a fresh individual has taken birth at the same position of the dead
   individual so that the population conservation is maintained.

%---------------------------------------------------------------------------
\begin{figure}[top]
\begin{center}
\includegraphics[width=6.5cm]{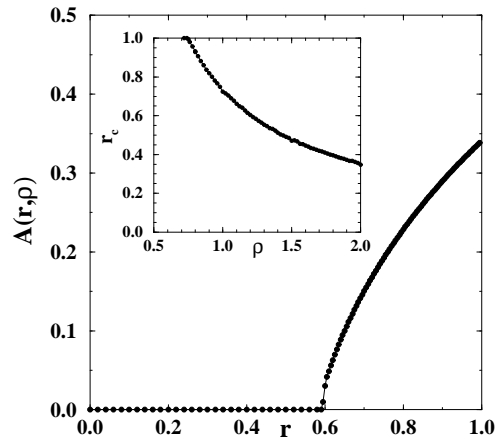}
\end{center}
\caption{
The variation of the activity $A(r)$ of the SIS model on DLFN
with the infection probability $r$ for $\rho=1.2$ and it vanishes at $r_c$. The inset shows
the variation of the critical infection probability $r_c(\rho)$ with
population density $\rho$.
}
\end{figure}
%---------------------------------------------------------------------------

      When an individual dies, the deletion of all his links may severely affect the
   distribution of distances between all pairs of nodes in the system. In
   fact it is expected that in general the distance between an arbitrary
   pair of nodes should increase due to the death of an individual, which 
   thus enhances the values of ${\cal D}(N,t)$ and ${\cal D}'(N,t)$. On the other hand 
   the newly born individual also start diffusing in the system and start 
   building up links of acquaintances with other individuals of the network. 
   Therefore the magnitudes of the diameters decrease again. As a result
   the net effect of the two competitive processes of simultaneous death 
   and birth of individuals is to make the diameters fluctuate around their 
   steady averages whose magnitudes must depend on the rate $q$ of
   death/birth processes. In Fig. 3 we show the time variation of these diameters and for
   $L=64, N=64$ the diameter ${\cal D}(q,N,L)$ has a value very near to 6 for $q$ = 0.43
   where as the ${\cal D}'(q,N,L)$  is around 2.85. As expected the diameters
   increase with decreasing $q$. 

      Finally we study a susceptible-infected-susceptible model on the 
   DLFN. At any time a lattice site may be occupied by a number of persons.
   If at least one of them is infected, each of the other healthy persons at that site
   has a probability $r$ to become infected and with a probability $1-r$ it
   remains healthy. An infected person at time $t$ becomes healthy at the next time step.
   For a certain average density $\rho$, the average fraction of infected persons
   in the system fluctuates but maintains a steady time independent average $A(r,\rho)$. In Fig. 4
   we show that the average activity $A(r,\rho)$ vanishes for $r<r_c$ and
   it continuously increases beyond $r_c$. The threshold $r_c$ is 
   the critical point of a phase transition from a completely healthy society
   to an infected society. The $A(r,\rho)$ plays the role of the order parameter
   in this transition. We also notice that $r_c$ is in general a function of 
   the population density $\rho$. In the inset of Fig. 4 we plot the variation of
   $r_c(\rho)$ with $\rho$. The value of the critical infection probability
   decreases with increasing the population density i.e., more the density
   it is more likely that the infection really spreads. On the other hand,
   below a certain density $\rho < \rho_c$ infection does not spread at all
   even with the maximum possible infection probability $r_c=1$.
   For the square lattice we estimate $\rho_c \approx 0.75$. 

      A number of different aspects of this model may be of interest. On average a human
   being remains more or less localized up to his/her home, home city or home country. Therefore
   perhaps it would be better to consider their motion as sub-diffusive ($R^2(t) \sim t^{2\mu}$ with $\mu<1/2$)
   rather than normal diffusion. Secondly DLFN may be important to study the reaction
   kinetic networks of two species reversible or irreversible chemical reactions $A+B \leftrightarrow C$.
   
      To summarize, we have considered the evaluation of the mutual friendship network
   in a dynamic model of a society. Each member of the society executes a diffusive motion.
   Members of the society represent nodes of the network and their mutual friendships are
   the links. The dynamical random graph associated with the network shows a novel dimension dependent phase
   transition. With a certain death/birth probability the network displays the
   six degrees of separation effect. We also observe that such a society remains always healthy
   if the average population density is below certain threshold value, which should have very
   important practical consequences.

   I thank P. Sen for some initial discussions, D. Dhar for
   some useful comments and Brian Hayes for pointing out the
   paper by J. S. Kleinfeld \cite {Milgram}.


\begin{thebibliography}{90}
\bibitem {sublattice} Here is a point of detail. Since each person jumps to one of its 
         neighbouring lattice point at each time step, individuals randomly distributed on a 
         square lattice are divided into two interpenetrating sublattices and there is no
         friendship at all between these two socities. Therefore we drop individuals on one
         sublattice initially.
\end{thebibliography}
\end{document}